\documentclass[aps,prd,twocolumn,showpacs,nofootinbib,floatfix,superscriptaddress]{revtex4-1}

\usepackage{amssymb}
\usepackage{graphicx}
\usepackage{amsmath}
\usepackage{rotating}
\usepackage{dcolumn}
\usepackage{url}
\usepackage{graphicx,color}
\usepackage{lineno}
\usepackage{commath}
\usepackage{xspace}
\usepackage{amsfonts}
\usepackage{amsbsy}

\usepackage{tabularx}
\newcolumntype{R}{>{\raggedleft\arraybackslash}X}

\font\FermiSmallfont=cmssq8 scaled 1200

\def\UMDppthead#1#2#3{
\null 
\begin{center}\vskip -1.0truein{\hbox to 7.5truen {
\hfill 
\vbox to 1in {\vfill \FermiSmallfont
              \hbox{#1}
              \hbox{#2}
              \hbox{#3}
              \vfill}
}}\vskip-0.0truein\end{center}}

\def\apj{Astrophys.\ J.}

\def\prd{Phys. Rev. D}

\newcommand{\araa}{{Annu. Rev. Astron. Astrophys.\,}}
\newcommand{\cosmomc}{\textsc{CosmoMC}\xspace}
\newcommand{\gplink}{\url{https://github.com/dkirkby/gphist}\xspace}

\begin{document}

\title{Model independent inference of the expansion history and implications for the \\ growth of structure}

\author{Shahab Joudaki} 
\affiliation{Centre for Astrophysics \& Supercomputing, Swinburne University of Technology, P.O. Box 218, Hawthorn, VIC 3122, Australia}
\affiliation{ARC Centre of Excellence for All-sky Astrophysics (CAASTRO)}
\affiliation{Department of Physics, University of Oxford, Denys Wilkinson Building, Keble Road, Oxford OX1 3RH, U.K.}
\author{Manoj Kaplinghat} 
\affiliation{Department of Physics and Astronomy, University of California, Irvine, CA 92697, USA}
\author{Ryan Keeley} \email{Corresponding author: rkeeley@uci.edu}
\affiliation{Department of Physics and Astronomy, University of California, Irvine, CA 92697, USA}
\author{David Kirkby} 
\affiliation{Department of Physics and Astronomy, University of California, Irvine, CA 92697, USA}


\begin{abstract}
We model the expansion history of the Universe as a Gaussian process and find constraints on the dark energy density and its low-redshift evolution using distances inferred from the Luminous Red Galaxy (LRG) and Lyman-alpha (Ly$\alpha$) datasets of the Baryon Oscillation Spectroscopic Survey, supernova data from the Joint Light-curve Analysis (JLA) sample, Cosmic Microwave Background (CMB) data from the Planck satellite, and local measurement of the Hubble parameter from the Hubble Space Telescope ($\mathsf H0$). Our analysis shows that the CMB, LRG, Ly$\alpha$, and JLA data are consistent with each other and with a $\Lambda$CDM cosmology, but the ${\mathsf H0}$ data is inconsistent at moderate significance. Including the presence of dark radiation does not alleviate the ${\mathsf H0}$ tension in our analysis. While some of these results have been noted previously, the strength here lies in that we do not assume a particular cosmological model. We calculate the growth of the gravitational potential in General Relativity corresponding to these general expansion histories and show that they are well-approximated by $\Omega_{\rm m}^{0.55}$ given the current precision. We assess the prospects for upcoming surveys to measure deviations from $\Lambda$CDM using this model-independent approach.
\end{abstract}

\maketitle

\providecommand{\eqn}[1]{eqn.~(\ref{eqn:#1})}
\providecommand{\tab}[1]{Table~\ref{tab:#1}}
\providecommand{\fig}[1]{Figure~\ref{fig:#1}}
\providecommand{\zstar}{\ensuremath{z^\ast}}
\providecommand{\zmax}{\ensuremath{z_{\text{max}}}}
\providecommand{\sH}{\ensuremath{{\cal H}}}

\begin{section}{Introduction}

The $\Lambda$CDM model, with a cosmological constant ($\Lambda$), cold dark matter (CDM), and baryons, provides an excellent fit to cosmological observations at both low and high redshift \citep{planck15,alam16,Betoule:2014frx, Hildebrandt16}. However, as the statistical precision of datasets have improved, the standard $\Lambda$CDM model has increasingly pointed towards the existence of dataset discordances, most notably a 3.4$\sigma$ tension between the direct measurement of the Hubble constant \citep{Riess:2016jrr} and that inferred from cosmic microwave background (CMB) temperature measurements by Planck \citep{planck13,planck15}. Further moderate discordances include the Planck CMB temperature with the Lyman-$\alpha$ forest of the Baryon Oscillation Spectroscopic Survey (BOSS; \citep{Font-Ribera:2013wce,Delubac:2014aqe,Bautista:2017zgn}), Planck Sunyaev-Zel'dovich cluster counts \citep{sz2014,sz2016}, and weak gravitational lensing measurements by the Canada-France-Hawaii Telescope Lensing Survey (CFHTLenS; \cite{heymans13, joudaki16a}) and the Kilo Degree Survey (KiDS; \cite{Hildebrandt16,joudaki16b,kohlinger17,Joudaki:2017zdt,uitert17}).

The tensions among datasets could be due to underestimated systematic effects associated with the experiments, or it may point to physics beyond the standard $\Lambda$CDM cosmology (e.g.~\cite{Canac:2016smv,joudaki16b,addison17}).  Examples of physics beyond the standard cosmological model include a time evolving equation of state for the dark energy fluid (e.g.~\cite{cds98,Zhao:2017cud}), an infrared modification to General Relativity (GR;~e.g.~\cite{Dvali:2000hr,Carroll:2003wy}), or a coupling of matter and dark energy (e.g.~\cite{freese87, Amendola:1999er}). An approach independent (as far as possible) of a cosmological model could be very useful given the lack of concrete directions to understand the larger cosmological constant problem (e.g.~\citep{weinberg89, ct92, fth08}).

Motivated by these observations, we test the $\Lambda$CDM model by inferring the expansion history and growth of structure in a ``model-independent'' manner using the method of Gaussian processes (GP; e.g.~\cite{Rasmussen:2006xyz}).  GP regression is compelling since it is both more flexible and more data driven than parametric approaches \cite{Ivezic:2014xyz}.  Performing such a regression analysis with GP is additionally useful since it avoids the problem of over fitting which is ubiquitous for polynomial regression.

A model independent approach runs into two issues: what freedom do we allow at the redshift of last scattering and how do we include the data on the growth of structure in a model-independent manner? We adopt a compromise in this work by assuming that at the time of last scattering the Universe can be described by a model based on General Relativity with dark matter, baryons, photons, three active neutrinos, and possibly extra relativistic degrees of freedom. We compute the growth history in a model-independent manner from the expansion history with the assumption of General Relativity and then compare it to observations. 

Previous studies have used GP regression to study the expansion history generally, and to study the dark energy equation of state specifically.  An early example of the former is the study by Shafieloo, Kim, and Linder~(2012) \cite{2012PhRvD..85l3530S}, where the authors generate a GP for $H(z)^{-1}$ (without dividing out a fiducial model) and use the regression on the Union 2.1 dataset of supernovae (SNe). They take the results of their GP regression, derive a posterior for the deceleration parameter $q = -a \ddot{a}/\dot{a}^2$, and find agreement with the $\Lambda$CDM cosmology. Our method builds on these investigations and we also include additional datasets and distance measures ($D_{A}$).

An example of modeling the dark energy equation of state $w(z)$ using GP is found in Holsclaw~et~al.~\cite{2010PhRvD..82j3502H, 2010PhRvL.105x1302H, 2011PhRvD..84h3501H}. They generate the dark energy density from $w(z)$, and along with fiducial values for the matter density and radiation density, calculate a luminosity distance.  They use simulated datasets and the Constitution set of SNe to constrain $w(z)$.  They point out, however, that the reconstructed equation of state is sensitive to the assumed fiducial values.  The authors also discuss prospects for using the baryon acoustic oscillation (BAO) feature and CMB to reconstruct the equation of state.

A recent analysis parameterized the late-time expansion rate $H(z)$ with cubic splines and discussed the tension between the local measurement of $H_0$ and its Planck inference~\cite{Bernal:2016gxb}. They pointed out that the $H_0$ tension could be pointing to a smaller sound horizon at the drag epoch ($r_{\rm drag}$), since they are both derived parameters of the expansion rate. Using the temperature and low-$\ell$ polarization data, they concluded that including extra radiation at recombination can relieve this tension.  However, the inclusion of the high-$\ell$ polarization data disfavors this interpretation. We will compare to these results in the discussion of the $H_0$ tension in the forthcoming sections.

In Section~\ref{sec:data}, we outline the different datasets used in our analysis, which span a wide range of redshifts from the present to the epoch of recombination. In Section~\ref{sec:expansion}, we describe the setup for our GP regression and the inferences on the expansion history that the regression provides, both with present data and forecasted with DESI. In Section~\ref{sec:potential}, we use the GP results to infer the growth history and dark energy density with redshift. We conclude with a summary of our results in Section~\ref{sec:conclusion}.

\end{section}

\begin{section}{Data}\label{sec:data}

We include low-redshift distances from Beutler~et~al. (2016) \cite{Beutler:2016ixs}, who analyzed the clustering of more than a million galaxies in the redshift range $0.2<z<0.75$ from the final BOSS data release (DR12) to extract the baryon acoustic oscillation signal. The angular diameter distance, $D_{A}(z)$, and Hubble parameter, $H(z)$, are measured at the 1 -- 3\% level in three redshift bins centered at $z = 0.38$, $0.51$, and $0.61$. We label this dataset `LRG'.

From Bautista et al.~(2017) \cite{Bautista:2017zgn}, we obtain high-redshift distances calculated from the BAO feature in the flux correlation function of the Ly$\alpha$ forest. Bautista et al.~(2017) use more than 150,000 quasars in the redshift range $2.1 \leq z \leq 3.5$ from DR12 of the BOSS SDSS-III. They measure the Hubble distance and the angular diameter distance with respect to the size of the sound horizon at the drag epoch, $r_{\rm drag}$, at an effective redshift of $z=2.33$.  This dataset is labeled `Ly$\alpha$' in the rest of this paper.

New results by the BOSS collaboration on the Ly$\alpha$--quasar correlation function at $z=2.4$ have just been released~\cite{Bourboux:2017cbm}. The results are consistent with the Planck cosmology~\cite{planck15} at the 2-$\sigma$ level. The small deviation from Planck $\Lambda$CDM is, however, difficult to model because the inferred $D_H$ at $z=2.4$ is larger, while $D_A$ is smaller. When combined with the auto-correlation data~\cite{Bautista:2017zgn}, the Planck cosmology shows a moderate 2.3-$\sigma$ tension~\cite{Bourboux:2017cbm}. This finding is consistent with the previous BOSS Ly$\alpha$ results~\cite{Delubac:2014aqe}. A detailed discussion of this moderate tension in the Lyman-$\alpha$ and Planck datasets in terms of alternate models found no satisfactory solution~\cite{Aubourg:2014yra}. In particular, alternative solutions could not simultaneously fit all the BAO data and hence were not preferred over the flat $\Lambda$CDM model. If the tension becomes stronger, it would be interesting to use our model-independent method to search for a possible solution. For the present, we do not include this new cross-correlation dataset, or the older dataset, in our analysis.

For our baseline results, we consider a fiducial value for $r_{\rm drag} = 147.36$ Mpc for the BAO measurements.  Since the relative uncertainty in $r_{\rm drag}$ is significantly smaller than the uncertainty in the measured value of the ratios $D_{H} / r_{\rm drag}$ and $D_{A} / r_{\rm drag}$ from the LRG and Ly$\alpha$ datasets, we take the uncertainties in $D_{H}$ and $D_{A}$ to arise only from the uncertainty in the ratios for the main analysis. We relax this assumption when discussing an expanded parameter space where the error on $r_{\rm drag}$ becomes comparable to the BAO measurement errors.

We include the direct measurement of the Hubble constant by Riess~et~al.~(2016) \cite{Riess:2016jrr}, who used the Wide Field Camera 3 on the Hubble Space Telescope to observe Cepheid variables in the same host galaxies as recent Type Ia supernovae to anchor its $z=0$ magnitude-redshift relation. Riess~et~al.~(2016) determined the distances to low-redshift anchors such as the megamaser system NGC 4258 and the Large Magellanic Cloud more robustly, and increased the number of observed local Cepheids in regions such as M31, the Large Magellanic Cloud, and the Milky Way. These improvements led to the estimate $H_0 = 73.24 \pm 1.74 \,{\rm km} \, {\rm s}^{-1} \, {\rm Mpc}^{-1}$, which we refer to as `${\mathsf H0}$'.

For luminosity distances inferred using Type Ia supernovae, we include the binned supernovae from Betoule et al.~(2013, 2014) \cite{Betoule13, Betoule:2014frx}. The 740 SNe of the SDSS-II and SNLS collaborations (joint light-curve analysis sample) are compressed into 31 redshift bins between $0.01<z<1.3$. These constraints are effectively on the ratio $D_{L} / D_{{H}_0}$, so we marginalize over the normalization of this distance modulus. We denote this dataset `SN'.

We consider the CMB temperature and polarization data from the Planck satellite \citep{planckone2015,planck15} to derive posteriors for the Hubble distance and angular diameter distance to the redshift of last scattering, $z_*$. The Planck dataset includes TT, EE, TE, and lowP angular power spectra. We refer to this dataset as `CMB'.  In Section~\ref{cmbc}, we discuss the key physics that controls the measured covariance matrix of $D_{H}$ and $D_{A}$ at the last scattering surface. Fiducially, we do not include the power spectrum of the CMB lensing potential ($\phi$) as part of the CMB dataset to avoid mixing high-$z$ and low-$z$ measurements (as the lensing kernel peaks at low redshift \cite{2012PhRvD..86b3526J, Ade:2015xua}).

Another epoch that lends itself to a model-independent analysis is Big Bang Nucleosynthesis where constraints on the expansion history have been obtained independent of a cosmological model~\cite{Carroll:2001bv}. We do not include it here given the many e-folds of scale factor between last scattering and the epoch when light elements form.

\begin{subsection}{Understanding the CMB constraint}
\label{cmbc}

The angular size of the sound horizon is given by the radius of the sound horizon at last scattering, $r_{s}$, divided by the angular diameter distance $D_A$ to last scattering: $\theta_{s} = \frac{r_{s}}{D_{A}(z_*)}$. The radius of the sound horizon is $r_{s} = \int^\infty_{z_*} D_{H}(z) c_{s}(z)/c~dz$, which scales with $D_H(z_*)$. Here, $c_s/c$ is the sound speed relative to the speed of light, and $D_H$ is the Hubble distance. We will discuss the impact of new physics on $r_s$ in Section~\ref{sec:expanded}.

With only information about the angular size of the sound horizon $\theta_{s} \propto D_{H}(z_*)/D_{A}(z_*)$, the $D_{H}(z_*)$ and $D_{A}(z_*)$ measurements would be fully degenerate. This degeneracy is broken by measuring the wavenumber related to photon diffusion, $k_{D}$. Diffusion is a random walk, so the diffusion length ($\propto 1/k_{D}$) scales as the square root of the number of scatterings multiplied by the mean free path.  The number of scatterings is proportional to $D_H(z_*)$, which gives $k_{D} \propto 1/\sqrt{D_{H}(z_*)}$. Note that the effect of damping on the heights of the peaks is dictated by the quantity $k_{D}r_{s} \propto \sqrt{D_{H}(z_*)}$, which is independent of low-redshift physics \cite{Hou:2011ec}.

Given constraints on $\theta_{s}$ and $k_{D}$, and knowing how they depend on $D_{H}$ and $D_{A}$, we can express the joint CMB log-likelihood for $D_H(z_*)$ and $D_A(z_*)$ in the following manner:
\begin{multline}
- 2 \log \mathcal{L} (D_{H}, D_{A})  \propto \\
\left( \frac{ k_1 D_{H}/D_{A} - \bar{\theta} }{\sigma_\theta} \right)^2 + \left( \frac{k_2/\sqrt{D_{H}}  - \bar{k}_{D}  }{\sigma_k} \right)^2,
\end{multline}
where $k_1$ and $k_2$ are constants, the bars represent the measured values, and $\sigma$ with a subscript is the uncertainty in the measured value corresponding to the subscript. We use this likelihood function to derive the covariance matrix, $C$, as the inverse of the Fisher matrix, $\mathcal{F}_{ij} = \langle \frac{d^2 \log \mathcal{L}}{d x_i d x_j} \rangle$, where $x_{i,j}$ are any generic parameters:
\begin{equation}
C = \begin{bmatrix}
4 D_{{H},0}^2 \left( \frac{\sigma_k}{\bar{k}_{D}} \right)^2 & 4 D_{{A},0} D_{{H},0} \left( \frac{\sigma_k}{\bar{k}_{D}} \right)^2 \\
4 D_{{A},0} D_{{H},0} \left( \frac{\sigma_k}{\bar{k}_{D}} \right)^2 &   D_{{A},0}^2 \left[ 4\left( \frac{\sigma_k}{\bar{k}_{D}} \right)^2  + \left( \frac{\sigma_\theta}{\bar{\theta}} \right)^2\right]\,
\end{bmatrix} .
\end{equation}
Using the values from our simplest Markov Chain Monte Carlo case (MCMC; using \cosmomc \cite{Lewis:2002ah}), `$\Lambda$CDM: TT', where $100\theta_{s} = 1.04131 \pm0.00051$, $k_{D} = 0.14049 \pm 0.00053$~Mpc$^{-1}$, $D_H = 1.927 \times 10^{-1}$~Mpc, and $D_A = 1.275 \times 10^1$~Mpc, we calculate $C_{11} = 2.15 \times 10^{-6}$~Mpc$^2 , C_{12} = 1.42 \times 10^{-4}$~Mpc$^2$, and $C_{22} = 9.44 \times 10^{-3}$~Mpc$^2$.

Compared to the actual covariance matrix for that same \cosmomc run, where $C_{11} = 1.69 \times 10^{-6}$ Mpc$^2$, $C_{12} = 5.89 \times 10^{-5}$ Mpc$^2$, and $C_{22} = 2.24 \times 10^{-3}$ Mpc$^2$, our approximation overestimates the uncertainties of $D_{H}$ and $D_A$ by about a factor of 1.1 along the $D_H$ direction and by a factor of 2.1 along the $D_{A}$ direction. This approximation worsens as CMB polarization information is included in the \cosmomc calculation.  This is likely because adding more data like `lowP' or `TE+EE' brings in more information that constrains $D_H$ and $D_A$ indirectly without impacting $k_d$ and $\theta_s$. This is reflected in the covariance matrices; both the approximate and actual covariance matrices decrease with additional data, but the actual covariances decrease faster.

This exercise shows that other features in the CMB angular power spectrum (not just $\theta_{s}$ and $k_{D}$) constrain $D_{H}$ and $D_{A}$. Thus, it is important to examine how new physics at the last scattering surface can bias our inferred expansion history at late times. We discuss this in Section~\ref{sec:expanded}.

\end{subsection}
\end{section}

\begin{section}{Expansion history}
\label{sec:expansion}

\begin{figure}[t]
\includegraphics[width=\linewidth]{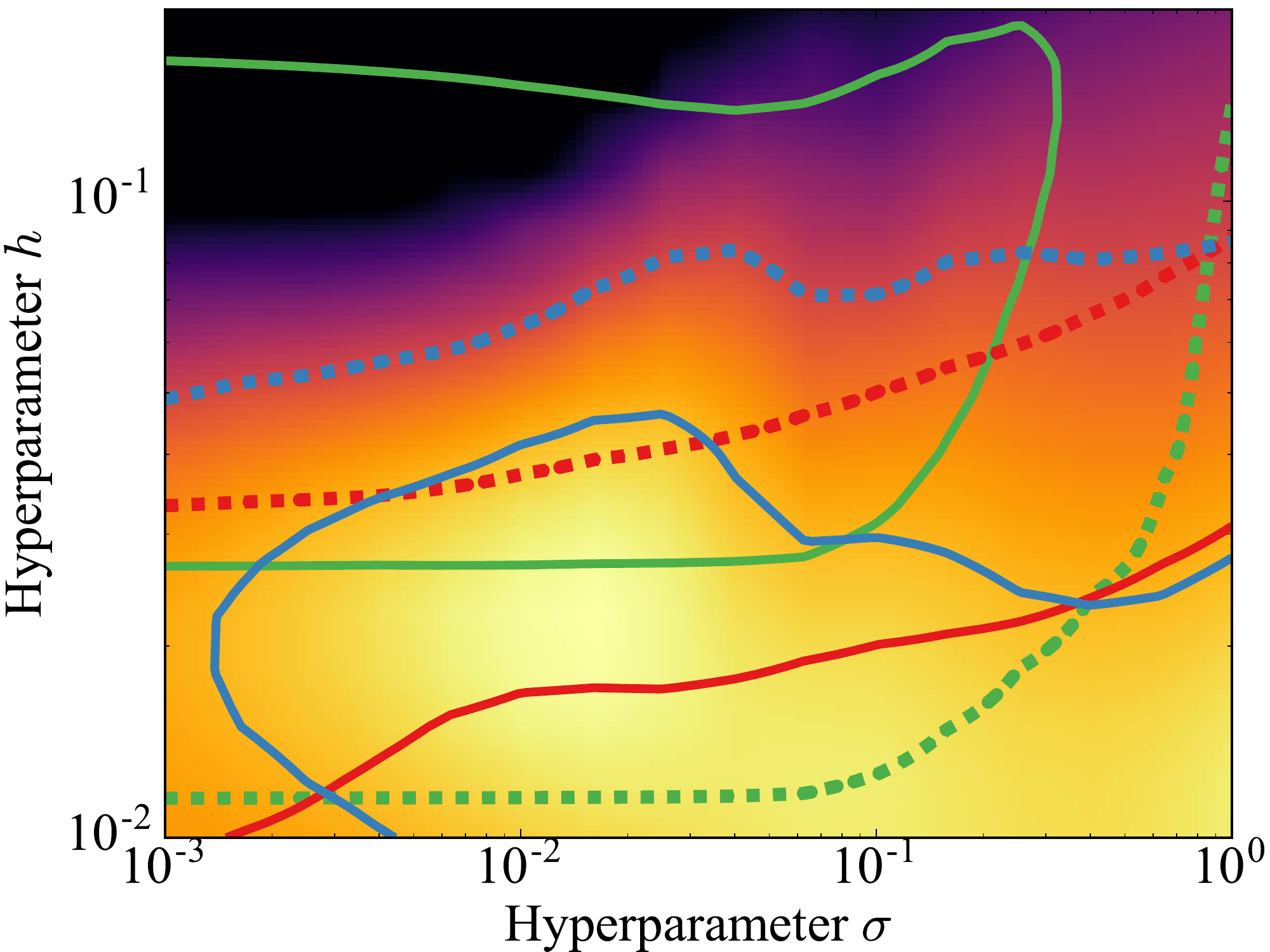}
\caption{\label{nuisance} Plotted are the 1 and 2-$\sigma$ contours of the posterior of the hyperparameters $\{h, \sigma\}$ that generate the GP. The solid lines correspond to the 1-$\sigma$ contours, and dashed lines the 2-$\sigma$ contours.  Green corresponds to the H0-Ly$\alpha$ combination, red the CMB-SN-LRG combination, and blue the full ${\mathsf H0}$-Ly$\alpha$-CMB-SN-LRG combination. The color map also corresponds to the full combination. It is apparent that the GP regression favors certain values of the hyperparameters.  Particularly, the CMB-SN-LRG, which is consistent with the fiducial model, does not meaningfully constrain $\sigma$, which describes the correlation length of the fluctuations, and prefers small values of $h$, which describes the size of the fluctuations.}
\end{figure}

\begin{figure*}[t]
\includegraphics[width=\textwidth]{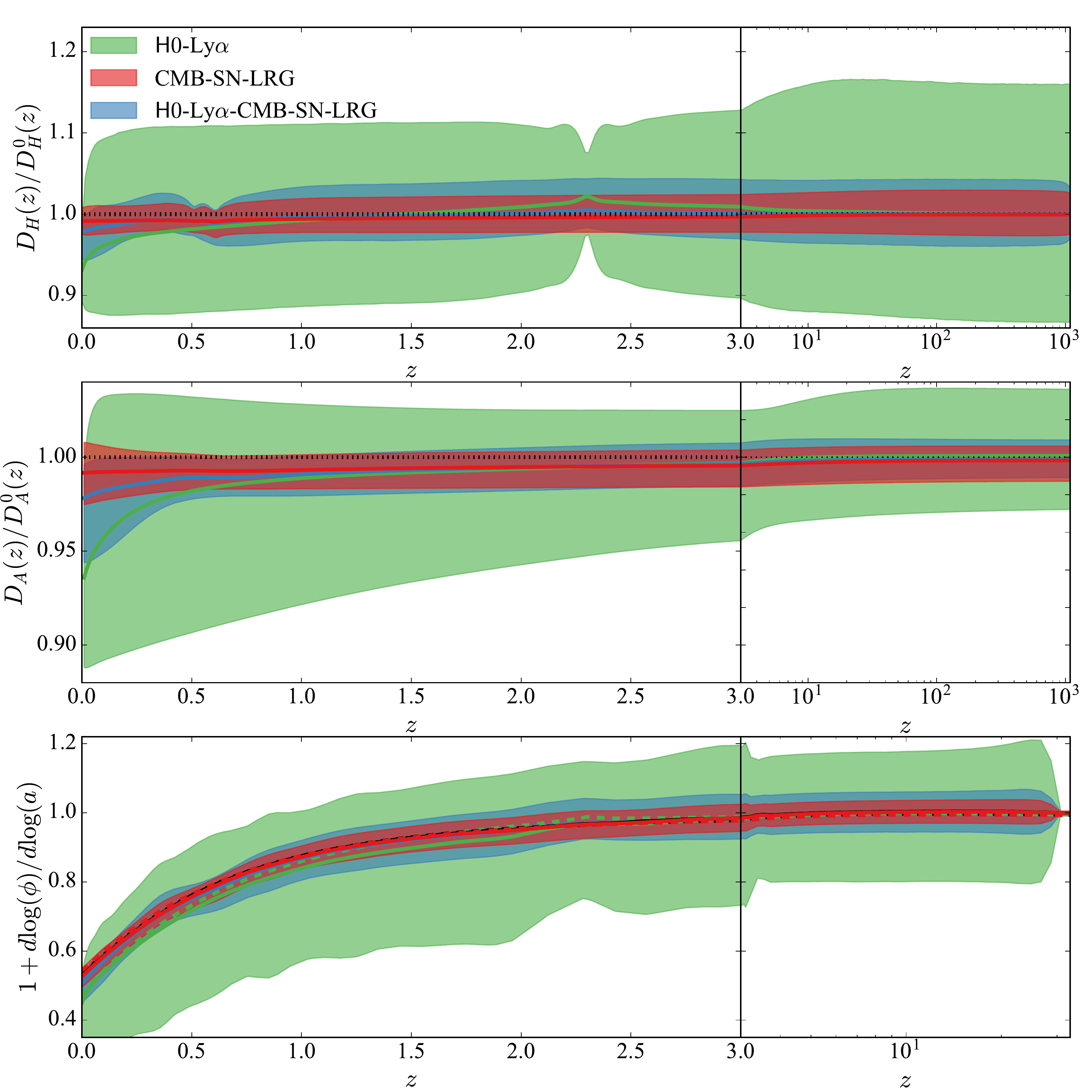}
\caption{\label{main} Expansion and growth histories determined by the GP regression for different combinations of the ${\mathsf H0}$, Ly$\alpha$, CMB, SN, and LRG datasets. The top panel shows the Hubble distances and the middle panel shows the angular diameter distances. These distances are plotted relative to those from the fiducial Planck $\Lambda$CDM cosmology. The shaded regions are bounded by the 90\% confidence level contours generated from the posterior probability for the quantity at that redshift. The solid lines denote the median values, and the dotted lines denote $\Lambda$CDM. As in Figure \ref{nuisance}, we show the combination that includes all of the different datasets, ${\mathsf H0}$-Ly$\alpha$-CMB-SN-LRG (blue), and the partition of the full dataset into combinations that are consistent with fiducial $\Lambda$CDM (CMB-SN-LRG, in red), and combinations that show moderate tension (${\mathsf H0}$-Ly$\alpha$, in green). The bottom panel shows the growth rate $f(z) = 1 + d \log(\phi) / d \log(a)$ derived from the expansion history. The dashed lines are $\Omega_{\rm m}^\gamma (z)$, where $\gamma$ is determined by the value that minimizes the squared distance between $f$ and $\Omega_{\rm m}^\gamma (z)$ weighted by the size of the uncertainty in $f(z)$.  $\gamma = 0.52, 0.53, 0.56$ for the ${\mathsf H0}$-Ly$\alpha$-CMB-SN-LRG, ${\mathsf H0}$-Ly$\alpha$, CMB-SN-LRG combinations, respectively.}
\end{figure*}

Assuming flatness, we constrain the expansion history $H(z)$ as a function of redshift using the Hubble distance $D_{H}(z)\equiv c/H(z)$, the angular diameter distance $D_{A}(z)\equiv D_{C}(z)/(1+z)$, and the luminosity distance $D_{L}(z)\equiv D_{C}(z)(1+z)$, where $D_{C}(z) = \int_0^z D_{H}(z') dz'$ is the comoving distance. We factor out a reference history $D_{H}^0(z)$, and model
\begin{equation}\label{eq:gamma}
\gamma(z) = \ln(D_{H}(z)/D_{H}^0(z))\,,
\end{equation}
as a GP with zero mean $\langle \gamma(z)\rangle = 0$
and a covariance function 
\begin{equation}
\langle \gamma(z_1) \gamma(z_2) \rangle =h^2 \exp(-(s(z_1)-s(z_2))^2/(2 \sigma^2))\,,
\end{equation}
with hyperparameters $h$ and $\sigma$ (e.g.~\cite{Rasmussen:2006xyz,Seikel:2013fda}). Note that we could expand our analysis in a simple way to non-flat cosmologies by including the curvature as an additional hyperparameter.  

We use the Planck+WP best fit to flat $\Lambda$CDM from Ade et al.~(2013) \cite{planck13} to calculate the reference history $D_{H}^0(z)$. Specifically, the fiducial model is constructed with Hubble constant $H_0 = 67.04~{\rm km}~{\rm s}^{-1}~{\rm Mpc}^{-1}$, present matter density $\Omega_{\rm m} = 0.3169$, present dark energy density $\Omega_{\rm DE} = 0.6831$, effective number of neutrinos $N_{\rm eff} = 3.046$, and one neutrino species with mass $m_\nu = 0.06$ eV.  The evolution variable $s(z)$ is taken to be
\begin{equation}
s(z) = \log(1+z)/\log(1+\zmax) \,,
\end{equation}
where $\zmax = 1090.48$, which matches the redshift of last scattering for the Planck+WP best fit. Note that $s(z)$ goes from 0 to 1 as $z$ changes from 0 to $\zmax$. We discretize $D_{H}(z)$ on a grid in $z$ and linearly interpolate $D_{H}(z)$ in $s(z)$ to obtain $D_{C}(z)$ through the following quadrature,
\begin{eqnarray}
&&D_{C}(z_{i+1}) = D_{C}(z_i) +D_{H}(z_i)(z_{i+1} - z_i)\nonumber\\
&&+ \frac{D_{H}(z_{i+1})- D_{H}(z_i)}{s(z_{i+1})-s(z_i)} \int_{z_i}^{z_{i+1}} \left(s(z)-s(z_i)\right) dz\,.
\end{eqnarray}
We use a fine enough grid in $z$ so that the errors from this quadrature are small.

GP regression is particularly useful since the regression occurs in an infinite-dimensional function space without overfitting. The covariance function of a GP corresponds to a Bayesian regression with an infinite number of basis functions \cite{Rasmussen:2006xyz}. GP regression works by generating a large sample of functions ($\gamma(z)$) determined by the covariance function.  These functions generated by the GP are transformed into Hubble distances and angular diameter distances, as in Eqn.~\ref{eq:gamma}.  Each of these generated expansion histories are given a weight determined by the likelihood of the data. These weighted expansion histories are then histogrammed at various redshifts in the range $0<z<1090.48$. Although standard libraries are available for GP, this application required custom code to support flexible constraints in the coupled $D_C(z)$ and $D_H(z)$ evolutions.  This code \cite{gphistdoi} is publicly available at \gplink.\footnote{After writing this paper we learned of a software package~\cite{2013ascl.soft03027S,Seikel:2012uu} with similar capabilities to ours.}

The results of the GP regression of course depend on the hyperparameters that determine the GP's covariance function.  Accordingly, we marginalize over these hyperparameters on a grid with values $0.01<h<0.2$ and $0.001<\sigma<1.0$. We calculate the posteriors of these hyperparameters (Figure~\ref{nuisance}) and find that they are well constrained when multiple datasets are used.

\begin{figure*}[thb]
\includegraphics[width=\textwidth]{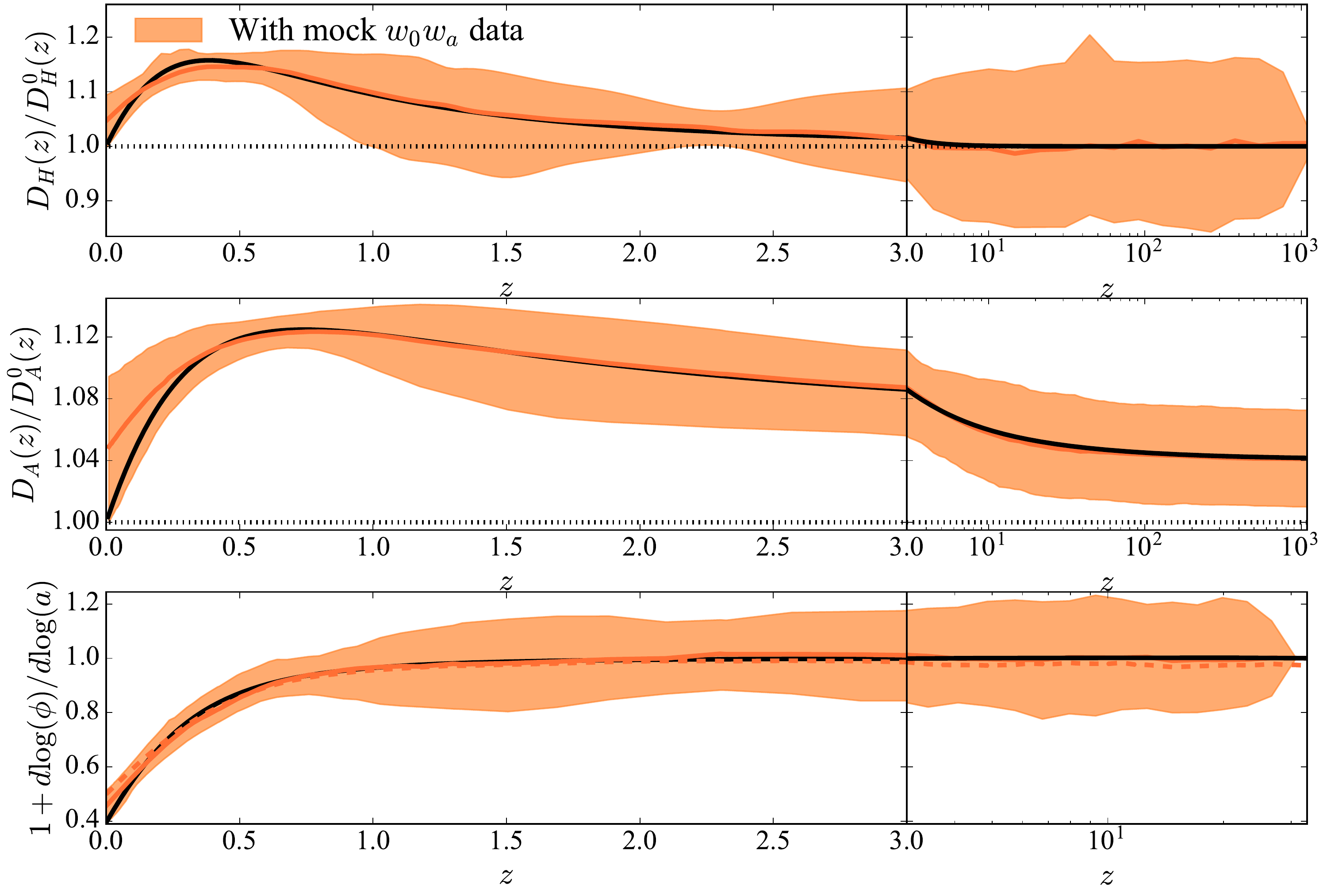}
\vspace{-0.7cm}
\caption{\label{bias} Results of the GP regression for mock data generated from a $w_0 w_a$ cosmology and resampled with the covariance matrices of the actual data. The shaded regions represent the posterior probabilities for the expansion and growth histories with redshift, bounded at 90\% CL. The results of the GP regression are in orange (solid orange lines denoting the medians), while the solid black lines correspond to the $w_0 w_a$ input cosmology. As in Figure~\ref{main}, the top panel shows the Hubble distance divided by the fiducial distance from a Planck $\Lambda$CDM cosmology, the middle panel shows the angular diameter distance, and the bottom panel shows the growth rate. The dashed orange line in the bottom panel is $\Omega_{\rm m}^\gamma (z)$, where $\gamma = 0.65$.}
\end{figure*}

The effects of the hyperparameters, $h$ and $\sigma$ can be understood in the following way. When looking at the prior distribution (i.e., inference without a dataset), the errors on the GP result vary proportionally with $h$ because it controls the size of the fluctuations.  Should data at some redshift pick out a certain scale for $h$, then that scale will set the size of the error bars at redshifts unconstrained directly by the data.  Typically, if the data are consistent with fiducial values up to some fluctuations, the GP regression will pick out smaller values of $h$. Exactly how small is determined by the size of the error bars; larger values of $h$ will tend to scatter the expansion history beyond the error bars and smaller values of $h$ will be consistent with the data. Hence, if the data are close to the fiducial values, only an upper limit on $h$ may be inferred and the error bars on the GP result will be small, as is the case for the CMB-SN-LRG case (see Figure \ref{nuisance}). The other hyperparameter, $\sigma$, controls the correlation length of the GP: if $\sigma$ is too large, the median of the regression misses a significant portion of the variance; if $\sigma$ is too small, the median of the regression overfits the data. If the constraints are sufficiently close to the fiducial values, or if the data prefers small values of $h$, the GP regression will not be able to constrain the values of $\sigma$.

In Figure~\ref{main}, we show the median expansion history and 90\% confidence level (CL) contours derived from the GP. The bands that overlap with the black dashed line at a given redshift are consistent with the best-fit Planck $\Lambda$CDM cosmology at that redshift (at 90\% CL). Similarly, bands that overlap with one another at a redshift represent datasets consistent with one another at that redshift. Globally, the mutual consistency between the CMB-Ly$\alpha$-SN-LRG datasets and the $\Lambda$CDM cosmology can further be seen from the fact that the median is relatively featureless and has tight error bars. However, the $H_0$ measurement is not consistent with the other datasets.  Note that the $D_{A}$ and $D_{H}$ medians are pulled to lower values and at $z=0$ there is only a small overlap between the ${\mathsf H0}$-Ly$\alpha$ dataset and the CMB-SN-LRG dataset.

A less obvious indicator of this inconsistency is seen in the relative size of the error bars for the CMB-SN-LRG data combination and the ${\mathsf H0}$-Ly$\alpha$-CMB-SN-LRG combination. If additional datasets are consistent with previous datasets, one would expect the GP from the union of the datasets to produce smaller error bars at all redshifts.  This is not the case with the inclusion of the ${\mathsf H0}$ dataset, implying some tension. Since the ${\mathsf H0}$ dataset is trying to pull $D_{H}$ below the fiducial value, the GP with the ${\mathsf H0}$ combination favors larger values of the hyperparameter $h$, which controls the scale of the fluctuations of the GP regression, than without it. This in turn produces larger error bars at all redshifts.

In other words, the exact precision of the GP constraints is sensitive to the concordance between the datasets included in the analysis. Beyond the discrepancies for $z \simeq 0$, the full data combination (${\mathsf H0}$-Ly$\alpha$-CMB-SN-LRG) constrains the expansion history (both in $D_H$ and $D_A$) to be consistent with the Planck $\Lambda$CDM cosmology at the 2\% level up to the redshift of last scattering. It is worth noting that when the size of the relative errors on $D_H$ are constant in redshift, the size of the relative errors on $D_A$ tends to decrease with redshift. This is because $D_A$ is the integral of $D_H$, and can be interpreted as the sum of $N$ independent random $D_H(z_i)$ variables. As a result, the error on $D_A$ grows as $\sqrt{N}$, while $D_A$ grows as $N$, with the relative error decreasing as $1/\sqrt{N}$. This explains why, despite having no CMB constraint, the ${\mathsf H0}$-Ly$\alpha$ dataset constraints on $D_A$ are tightest at high redshifts.

\newcommand{\cH}{{\cal H}}
\newcommand{\css}{c_{\textrm s}^2}

\begin{table*}[t]
\centering
\def\arraystretch{1.4}
\begin{tabular*}{\textwidth}{c @{\extracolsep{\fill}} ccccc}
 \toprule
 Cosmology  & Data & $D_H$ [Mpc] & $D_A$ [Mpc] & Correlation\\ \hline
 $\Lambda$CDM & TT & $(1.928 \pm 0.012) \times 10^{-1}$ & $(1.275 \pm 0.005) \times 10^1$ & 0.964 \\ 
  & TT+lowP & $(1.920 \pm 0.011) \times 10^{-1}$ & $(1.273 \pm 0.004) \times 10^1$ & 0.936 \\ 
  &  TT+lowP+lensing & ($1.926 \pm 0.010) \times 10^{-1}$ & $(1.276 \pm 0.004) \times 10^1$ & 0.952 \\
  & TT+TE+EE+lowP  & $(1.919 \pm 0.007) \times 10^{-1}$  & $(1.273 \pm 0.003) \times 10^1$ & 0.940 \\ \hline
 $\Lambda$CDM + $N_{\rm eff}$ & TT & $(1.848 \pm 0.047) \times 10^{-1}$ & ($1.203 \pm 0.041) \times 10^1$ & 0.988 \\ 
  & TT+lowP & $(1.912 \pm 0.031) \times 10^{-1}$ & $(1.266 \pm 0.023) \times 10^1$ & 0.981 \\ 
  & TT+lowP+lensing & ($1.919 \pm 0.029) \times 10^{-1}$ & $(1.269 \pm 0.022) \times 10^1$ & 0.983 \\ 
  & TT+TE+EE+lowP & $(1.926 \pm 0.023) \times 10^{-1}$  & $(1.278 \pm 0.016) \times 10^1$ & 0.987 \\ \hline
  $\Lambda$CDM  & TT & $(1.820 \pm 0.072) \times 10^{-1}$ & $(1.118 \pm 0.057) \times 10^1$ & 0.969 \\
+ ${{\mathrm d}n_{\mathrm s} / {\mathrm d}\ln k}$  & TT+lowP & $(1.923 \pm 0.054) \times 10^{-1}$ & $(1.287 \pm 0.039) \times 10^1$ & 0.974 \\ 
 + $\sum m_\nu$ + $Y_p$ + $N_{\rm eff}$ & TT+lowP+lensing & ($1.913 \pm 0.053) \times 10^{-1}$ & $(1.275 \pm 0.037) \times 10^1$ & 0.973 \\ 
  & TT+TE+EE+lowP & $(1.952 \pm 0.036) \times 10^{-1}$ & $(1.297 \pm 0.024) \times 10^1$ & 0.990 \\ \toprule
\end{tabular*}
\caption{\label{CMB} Hubble distances and angular diameter distances evaluated at the redshift of last scattering, $z_* = 1090$, along with their uncertainties and correlation coefficient, for each of the considered CMB datasets and cosmologies.}
\end{table*}

\begin{subsection}{Validation}
\label{valsec}

We now show that our methodology is able to accurately infer non-standard cosmologies from mock data. Concretely, we consider a dark energy model with a time-evolving equation of state parameterized by $w(z) = w_0 + \frac{z}{1+z} w_a$ \citep{cp01,Linder:2002et}. We choose $\{w_0, w_a\} = \{-2, 1\}$ and keep the other parameters fixed to their fiducial values.  This cosmology is an interesting choice for validation since, for large redshifts, the equation of state is close to the $\Lambda$CDM value of $-1$, but begins to differ significantly at low redshifts. We use this cosmology to generate mock data and apply a GP regression on this data.  The central values for the mock data are taken from the $D_{H}(z)$ and $D_{A}(z)$ for our $w_0w_a$ cosmology, resampled by the covariances from each of the used datasets.

The results of this validation are shown in Fig.~\ref{bias}. The median of the GP regression is indeed not precisely the same as the input distances due to the resampling of the data.  Any discrepancy between the input cosmology (black line, Fig. \ref{bias}) and the median of the GP (orange line, Fig. \ref{bias}) has only a small significance. The general features, such as the hump in both $D_{H}$ and $D_{A}$ at low redshift, of the input cosmology are recovered. In addition to demonstrating the ability of our GP regression to reproduce non-standard cosmologies, this validation shows that our main results are not particularly sensitive to the choice of fiducial model (that we divide out) since the recovered cosmology is substantially different from the fiducial cosmology.

\end{subsection}

\begin{subsection}{Expanded parameter spaces}\label{sec:expanded}

We have considered expanded and contracted covariances for the CMB data (to simulate new physics), specifically, by scaling the elements of the covariance matrix by a factor of two. The late-time constraints were insensitive to such changes in the covariance. We also explicitly considered expanded parameter spaces. We generated posteriors for $D_{H}(z_*)$ and $D_{A}(z_*)$ using \cosmomc \cite{Lewis:2002ah} for three different model cases and four different data cases. The three model cases are $\Lambda$CDM, a minimal case where only the standard six parameters are varied ($\Omega_b h^2, \Omega_c h^2, \theta_{\rm MC}, \tau, n_s, \ln{(10^{10} A_{\mathrm s})}$), a case where $N_{\rm eff}$ is also varied, and an extended case where the running of the scalar spectral index ${{\mathrm d}n_{\mathrm s} / {\mathrm d}\ln k}$, sum of neutrino masses $\sum m_\nu$, and primordial helium abundance $Y_p$ are varied along with $N_{\rm eff}$. The different CMB data cases include different combinations of the temperature (TT), low-$\ell$ polarization (lowP), high-$\ell$ polarization (TE+EE), and lensing data. These results are listed in Table \ref{CMB}.

The additional parameters yield constraints on $D_{H}(z_*)$ and $D_{A}(z_*)$ with larger uncertainties relative to those from the base case. However, the inferred expansion history showed no significant deviations when using either the expanded or contracted covariance matrices. This is because there is no significant shift in the $D_{H}(z_*)$ and $D_{A}(z_*)$ values when the extra parameters are introduced, and changes to the median values are consistent with the expanded errors. This indicates that conclusions about late-time effects such as dark energy domination and the growth of structure are largely independent of the specific CMB constraint.

To isolate the effects of adding $N_{\rm eff}$, we further examined the `$\Lambda{\rm{CDM}} + N_{\rm eff}$' model separately. In particular, we focused on the `TT' dataset that allows for the largest freedom; for this case the inferred value of $H_0 = 80.5^{+6.7}_{-9.0}~{\rm km}~{\rm s}^{-1}~{\rm Mpc}^{-1}$. The error on the inferred $D_H(z_*)$ (see Table \ref{CMB}) and correspondingly on $r_{\rm drag}$ is also large in this case. This means that we need to propagate the changes in $r_{\rm drag}$ to the BAO distance measurements. To do so, we need a model for how $r_{\rm drag}$ varies with $D_H(z_*)$. To gain an understanding of the covariance between the cosmological variables when $N_{\rm eff}$ is varied, we perform the following exercise. We start with a $\Lambda$CDM model with $N_{\rm eff}=3.046$, increase $N_{\rm eff}$ and then discuss the changes to the cosmological parameters required to get the TT power spectrum back to the $\Lambda$CDM TT spectrum.

\begin{figure*}[t!]
\includegraphics[width=\textwidth]{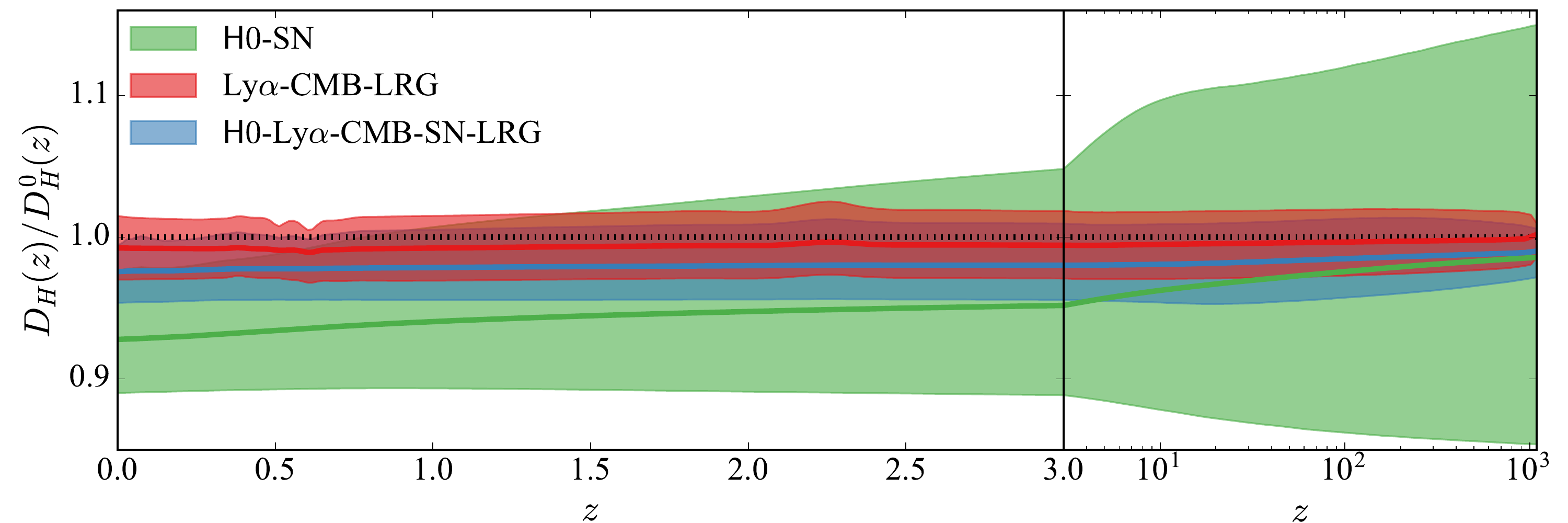}
\vspace{-0.7cm}
\caption{\label{rs_Neff} The GP results for $D_H$ considering the `$\Lambda$CDM + $N_{\rm eff}$' case, using only the `TT' constraint from the CMB. In green, we show the dataset combination that scales as a function of $D_H(0)$ (${\mathsf H0}$-SN). In red, we show the dataset combination that scales as a function of $D_H(z_*)$ (Ly$\alpha$-CMB-LRG); see discussion in text. As before, blue corresponds to the full dataset combination (${\mathsf H0}$-Ly$\alpha$-CMB-SN-LRG).
}
\end{figure*}

Increasing $N_{\rm eff}$ delays matter-radiation equality, i.e., decreases $z_{\rm eq}$ (redshift when matter and relativistic energy densities are equal). In order to obtain a good fit to the CMB data, we keep $z_{\rm eq}$ constant by increasing the physical matter density, $\Omega_{\rm m} h^2$. In addition, decreasing the baryon fraction by a small amount (keeping $\Omega_{\rm m} h^2$ fixed) to keep $k_Dr_s$ constant, one can maintain the same relative damping of the peaks as the $\Lambda$CDM model~\cite{Hou:2011ec}.

To keep the peak positions unchanged, we have to decrease $D_A(z_*)$ commensurate with the decrease in $r_s$ (so that $\theta_s$ doesn't change), which can be accomplished by increasing the dark energy density. This necessitates a decrease in $\Omega_{\rm m}$ to maintain a flat universe, which in turn requires an increase in $H_0$ to keep $\Omega_{\rm m} h^2$ (and hence $z_{\rm eq}$) unchanged. Using this model we find that the increase in $H_0$ is about 10\%, roughly consistent with what we find from the full MCMC for the TT case. This analysis shows in a simple way why an increase in $N_{\rm eff}$ is correlated with an increase in $H_0$ or vice-versa \cite{planck15,Canac:2016smv,Bernal:2016gxb}.

In addition to these changes, we found that an increase in the spectral index of the primordial power spectrum ($n_s$) leads to a better match. This is also evident in the contours plotted in Fig.~20 of Ref.~\cite{planck15}. With these changes and a small shift in the overall normalization (allowed by the uncertainty in the optical depth measurement), the changes to the spectrum from increasing $N_{\rm eff}$ can be made smaller than cosmic variance at $\ell < 2000$. We have checked this explicitly using the Python version of CAMB \cite{Lewis:1999bs}. 

Given this model, we can now predict the change to $r_{\rm drag}$. At fixed $z_{\rm eq}$, we have $D_H(z_*) \propto \sqrt{\Omega_{\rm m} h^2}$ (assuming $z_*$ changes are subdominant, which we verified). In addition, there is a correlated change in the baryon density $\Omega_b h^2$ and hence the sound speed, which implies that $r_{\rm drag}$ will not scale linearly with $D_H(z_*)$. For the model discussed above, we obtain $r_{\rm drag} \propto D_H(z_*)^{1.3}$. The MCMC results showed a steeper correlation: $r_{\rm drag} \propto D_H(z_*)^{1.5}$. The small discrepancy implies that we are not capturing all the available freedom in this simple model.

Given the above discussion, we generated expansion histories for the TT-only case using the model $r_{\rm drag} = r_{\rm drag, fid} \left(D_H(z_*)/D_H(z_*)_{\rm fid}\right)^{1.5}$. We scaled the BAO distances (both $D_A$ and $D_H$) by these $r_{\rm drag}$ values. This allows the large uncertainty in the CMB measurement for the TT-only case to impact the BAO measurements directly.  The results of the GP regression using this model is shown in Fig.~\ref{rs_Neff}. We did not find clear evidence that varying $N_{\rm eff}$ alleviates the $H_0$ tension.  Qualitatively, our results are in agreement with the findings of Ref.~\cite{Canac:2016smv} who discussed tension in cosmological datasets while allowing for the primordial power spectrum to be a knotted spline function.

It is worth noting that the $r_{\rm drag}$ values allowed by the TT data when $N_{\rm eff}$ is free to vary do change the low redshift $D_H(z)$ inferences, but the changes at $z=0$ are fairly minor. We ascribe this result to the fact that the decrease in $D_H$ is not significant enough (about 1-$\sigma$ given the expanded error) and that other low redshift measurements are consistent with Planck. We note that the CMB constraints in Table \ref{CMB} are consistent with the findings of Ref.~\cite{Bernal:2016gxb}. The key effect of varying $N_{\rm eff}$ is to enlarge the error on $D_H(z_*)$ (and hence $r_{\rm drag}$).  Adding polarization or lensing data reduces the error on $D_H(z_*)$ and pushes the median back to its value in $\Lambda$CDM.

\end{subsection}

\begin{subsection}{Forecasts}
\begin{figure*}[t!]
\includegraphics[width=\textwidth]{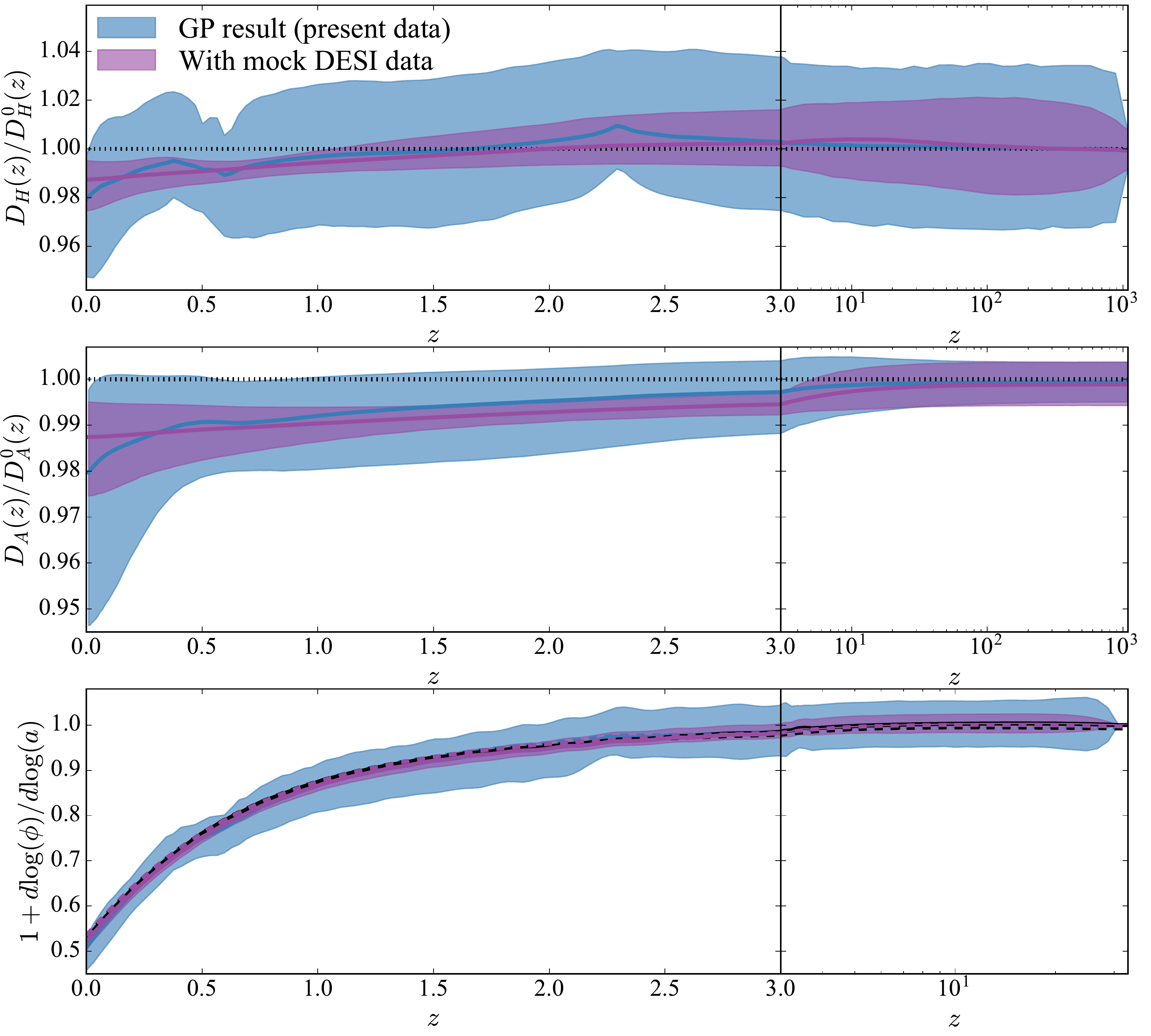}
\vspace{-0.84cm}
\caption{\label{DESI} The results of the GP regression, now with the inclusion of mock data from the upcoming DESI experiment. The~shaded regions correspond to the posterior probability of the expansion history with redshift, bounded at 90\% CL. The blue curves and shaded regions represent the results from the ${\mathsf H0}$-Ly$\alpha$-CMB-SN-LRG dataset. The purple curves and shaded regions further include mock DESI data generated from the median values of the ${\mathsf H0}$-Ly$\alpha$-CMB-SN-LRG regression. The black curves show the fiducial values for the quantities in each panel. The Hubble distances are in the top panel, angular diameter distances are in the middle panel, and growth rates are in the bottom panel. The dashed lines in the bottom panel show $\Omega_{\rm m}^\gamma (z)$ for $\gamma = 0.52, 0.54$ for the GP result and DESI forecast result, respectively.  The black dashed line shows the fiducial quantity $\Omega_{\rm m}^{0.55}(z)$.
}
\end{figure*}

We can use our methodology to forecast future constraints on the expansion history. As an example, we consider including, in addition to current data (${\mathsf H0}$-Ly$\alpha$-CMB-SN-LRG), information from the Dark Energy Spectroscopic Instrument (DESI, \cite{Aghamousa:2016zmz}). We use the projected uncertainties on $D_{H}$ and $D_{A}$ from the DESI Final Design Report \cite{DESI} and generate the central values from the median values of the GP regression with the ${\mathsf H0}$-Ly$\alpha$-CMB-SN-LRG dataset. Accordingly, DESI spans 24 redshifts bins between $0.65<z<3.55$, and has errors on $D_H$ and $D_A$ on the order of 1 -- 2\% for the lowest redshift bins and as large as 16\% for the highest redshift bin.

When the central values of the new data points are generated from the median result of the GP regression, the combination of the DESI data with the previous datasets yields a precision of $\lesssim 1\%$ across redshift, from the present to the last scattering surface. The tightest constraints in $D_H$ are located around $0.5<z<1.0$ while in $D_A$ they are around $1<z<4$, as seen in Figure~\ref{DESI}. In the event DESI follows the trend of current data, it will discern deviations from the fiducial $\Lambda$CDM cosmology at above 90\%~CL for all redshifts $z<1$ in $D_H$ and all redshifts $z<4$ in $D_A$. By contrast, without the DESI data, the only significant evidence for a deviation from $\Lambda$CDM occurs as a result of the ${\mathsf H0}$ dataset close to $z=0$.

It would also be possible to forecast the impact of future CMB or ${\mathsf H0}$ experiments. As discussed earlier, a factor of two reduction in the covariance matrix for the CMB does not particularly affect the low-redshift constraints. Improved uncertainties from a future ${\mathsf H0}$ experiment could lead to interesting new results, and we leave this for future work.

\end{subsection}

\end{section}

\begin{section}{Late-time growth of the gravitational potential}\label{sec:potential}

\newcommand{\cH}{{\cal H}}
\newcommand{\css}{c_{\textrm s}^2}

An avenue for looking for deviations from General Relativity on large scales is the correlated evolution of the late-time growth of the gravitational potential and the expansion history. We can use the space-space perturbed Einstein equations \cite{mb95} neglecting anisotropic stress and total pressure perturbation to write an equation for the gravitational potential:
\begin{equation}
\phi''+(4 + H'/H)\phi'+(3+2H'/H)\phi=0\,. \label{eq:growth}
\end{equation}
If we enter the well-known solution for the $\Lambda$CDM model, $\phi \propto (H/a)\int (aH)^{-3} da$, we find that $3(H^2)'+(H^2)''=0$, where primes denote derivatives with respect to $\ln(a)$. This is satisfied if the expansion rate is of the $\Lambda$CDM form: $H(z)^2 = c_1 (1+z)^3+c_2$ for constant $c_1$ and $c_2$.

Eqn.~\ref{eq:growth} can also be derived by starting with the assumption that the energy-momentum tensor is covariantly conserved and writing the perturbation equation for the total energy density at late times neglecting the radiation energy density and any anisotropic stresses. In addition, one needs to assume the hierarchy $k \gg \cH \gg k \css$ so as to be able to use the Poisson equation and neglect total pressure perturbations. This way of deriving Eqn.~\ref{eq:growth} may be useful in thinking about modified gravity theories where the Poisson equation is modified or the two gravitational potentials (typically labeled $\phi$ and $\psi$ \cite{mb95}) are not equal \cite{Bertschinger:2006aw,Daniel:2008et} but the energy-momentum tensor still satisfies the same conservation equations.

\begin{subsection}{Inferring the growth history}

We numerically solve Eqn.~\ref{eq:growth} for each generated expansion history in order to calculate the growth of the gravitational potential. We set the initial condition for this equation during the era of matter domination, specifically at $z=30$, which explains the narrowing of the contours of the growth history at that redshift. Choosing to set the initial condition at this redshift only requires the assumption that new physics is important solely at late times. We store both the gravitational potential ($\phi$) and its derivative encapsulated in the growth rate $f = 1 - d \ln (\phi)/ d\ln (1+z)$. As noted previously, the distance constraints determine a weight for each expansion history sampled by the GP.  The quantities $\phi$ and $f$, calculated for each sampled expansion history, are given this same weight, which allows us to calculate posteriors for $\phi$ and $f$.

A comparison of the growth function $D(a) = a\phi(a)$ to its measurement from the Dark Energy Survey (DES), South Pole Telescope (SPT), and Planck is shown in Fig.~\ref{growth}. These measurements are obtained by cross-correlating lensing maps of the CMB from Planck  and SPT with galaxy maps from DES~\cite{Giannantonio:2015ahz}. The errors on the growth function from galaxy clustering and galaxy-CMB lensing correlations are currently large, and broadly consistent with our inference from the expansion history. There is mild evidence that the measured growth function is systematically lower than the inferred one, and this provides an interesting target for future observations.

\newcommand{\dproplss}{D_\star}
\newcommand{\dprop}{D}

We have also investigated the redshift evolution of the growth rate. It is well known that, in GR, $f(z)$ can be accurately modeled by $\Omega_{\rm m}(z)^\gamma$ with $\gamma\simeq 0.55$ \citep{linder05,Ballesteros:2008qk}, where $\Omega_{\rm m}(z) \equiv \Omega_{\rm m}(0) (1+z)^3(H_0/H(z))^2$ is the matter density assuming the energy-momentum tensor of the matter component is separately conserved.

\begin{figure}[t!]
\includegraphics[width=\linewidth]{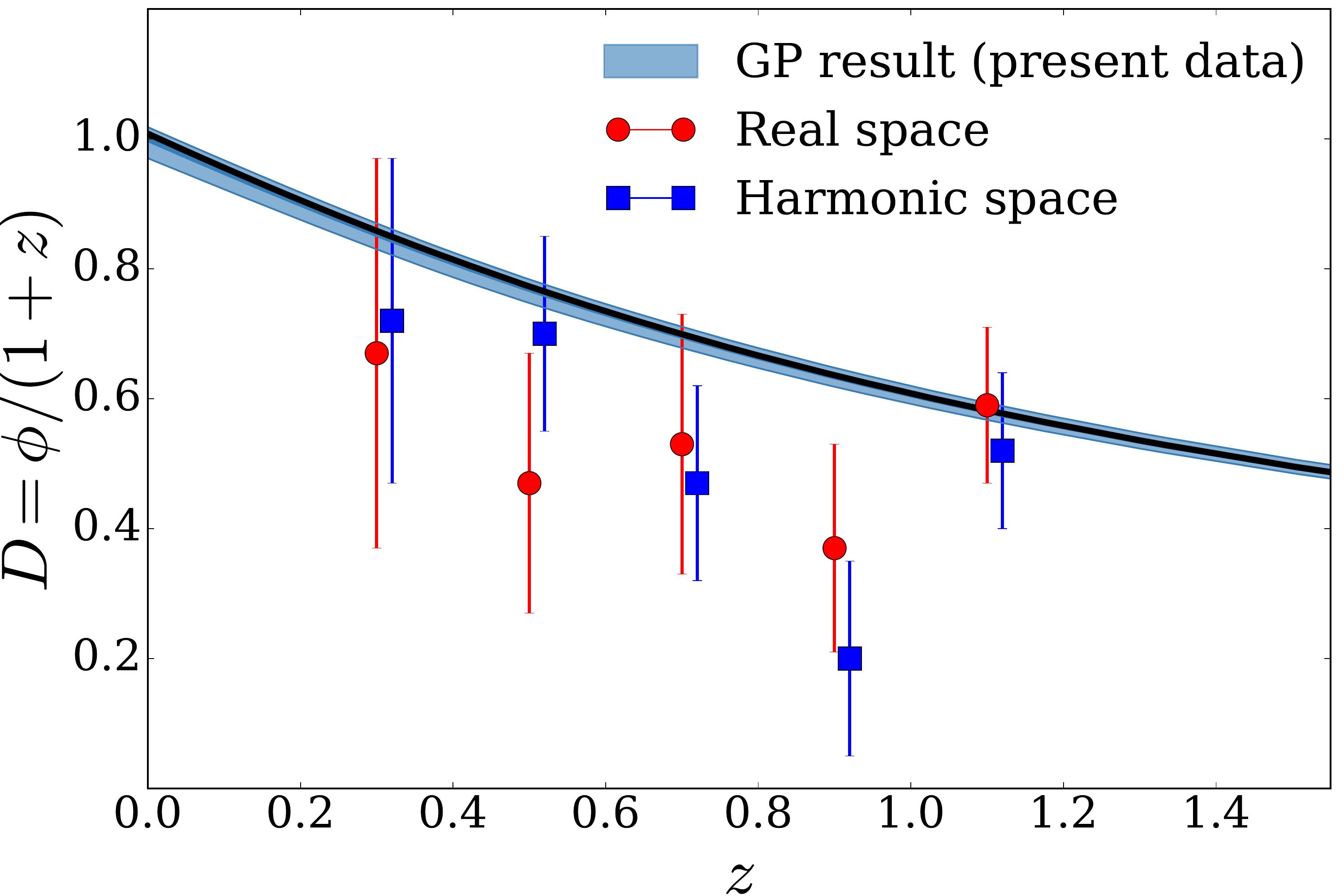}
\caption{\label{growth} The evolution of the gravitational potential multiplied by the scale factor where the median of the GP regression is normalized to unity at the present time and compared to galaxy clustering and galaxy--CMB lensing data from a joint analysis of DES, SPT, and Planck \cite{Giannantonio:2015ahz} (all shown at 68\%~CL). The GP result considers the full ${\mathsf H0}$-Ly$\alpha$-CMB-SN-LRG dataset (light blue). The black line corresponds to the fiducial $\Lambda$CDM cosmology. The difference between the red and blue points is that the former are calculated from an estimator of the growth function based off the two-point correlation function (real space), while the latter are calculated from a growth function estimator based off the angular power spectrum (harmonic space). Hence, they represent different techniques to calculate the same quantity.}
\end{figure}

In our model-independent method, $\Omega_{\rm m}(0)$ is not defined, so we start with the physical matter density at last scattering. Specifically, for each of the expansion histories generated by the GP, we calculate the value of the Hubble parameter at $z = z_*$, use that to obtain the total energy density at that redshift assuming GR, and then subtract off the energy density in radiation as defined by the fiducial model. We interpret the remaining quantity as the physical matter density at $z = z_*$. This physical matter density is then scaled to the critical density at other redshifts computed from the expansion histories as
\begin{equation}
\Omega_{\rm m} (z) = \left( \frac{3H^2(z_*)}{8 \pi G} -  \rho_{\rm r, fid} (z_*) \right) \left( \frac{1+z}{1+z_*}\right)^3  \frac{8 \pi G}{3 H^2(z)}.
\end{equation}
The benefit of this method is it maintains the model-independence at late times. The drawback is the need to make specific assumptions to obtain the matter density at last scattering. In particular, we need information about the energy density in relativistic degrees of freedom to determine the matter density. This can be done in a manner that is independent of late-time cosmology since the phase shift of the acoustic peaks \cite{Bashinsky:2003tk,Baumann:2015rya} and the damping tail \cite{Hu:1996mn,Hou:2011ec} in the CMB allow us to infer the energy density in non-interacting relativistic degrees of freedom \cite{Story:2012wx,Louis:2016ahn}. These observables in the angular power spectrum are determined by the evolution of the gravitational potential and expansion history at early times. In particular, the damping of the peaks is set by $k_{D} r_{s}$, which is manifestly independent of the low-redshift expansion history. The phase shift is proportional to $f_\nu \Delta \ell_{\rm peak}$, where $f_\nu$ is the fraction of energy density in non-interacting relativistic degrees of freedom (including standard model neutrinos) and $\Delta \ell_{\rm peak}$ is the spacing of peaks for modes that entered the horizon during radiation domination, first measured in the Planck 2013 data \cite{Baumann:2015rya,Follin:2015hya}. These arguments suggest that we can measure $N_{\rm eff}$ without degeneracy with the late-time expansion of the Universe.

Currently, there is no strong evidence for dark radiation. In computing $\Omega_{\rm m} (z)$, we use the standard cosmological radiation energy density (CMB photons and three massless active neutrinos). The constraints on $f(z)$ are shown in Figures~2 -- 4, and are consistent with the expectation that $f(z) = \Omega^\gamma_{\rm m} (z)$ with $\gamma = 0.55$, with a precision of 3 -- 4\% across redshift using the full combination of current data (${\mathsf H0}$-Ly$\alpha$-CMB-SN-LRG). This is not surprising since the expansion history is consistent with the fiducial history. Future surveys like DESI will have the power to substantially improve the constraints, reducing the uncertainty on the growth history to roughly 1\%, and
increasing the prospects for detecting deviations from the General Relativistic expectation.

\end{subsection}

\begin{figure}[t!]
\includegraphics[width=\linewidth]{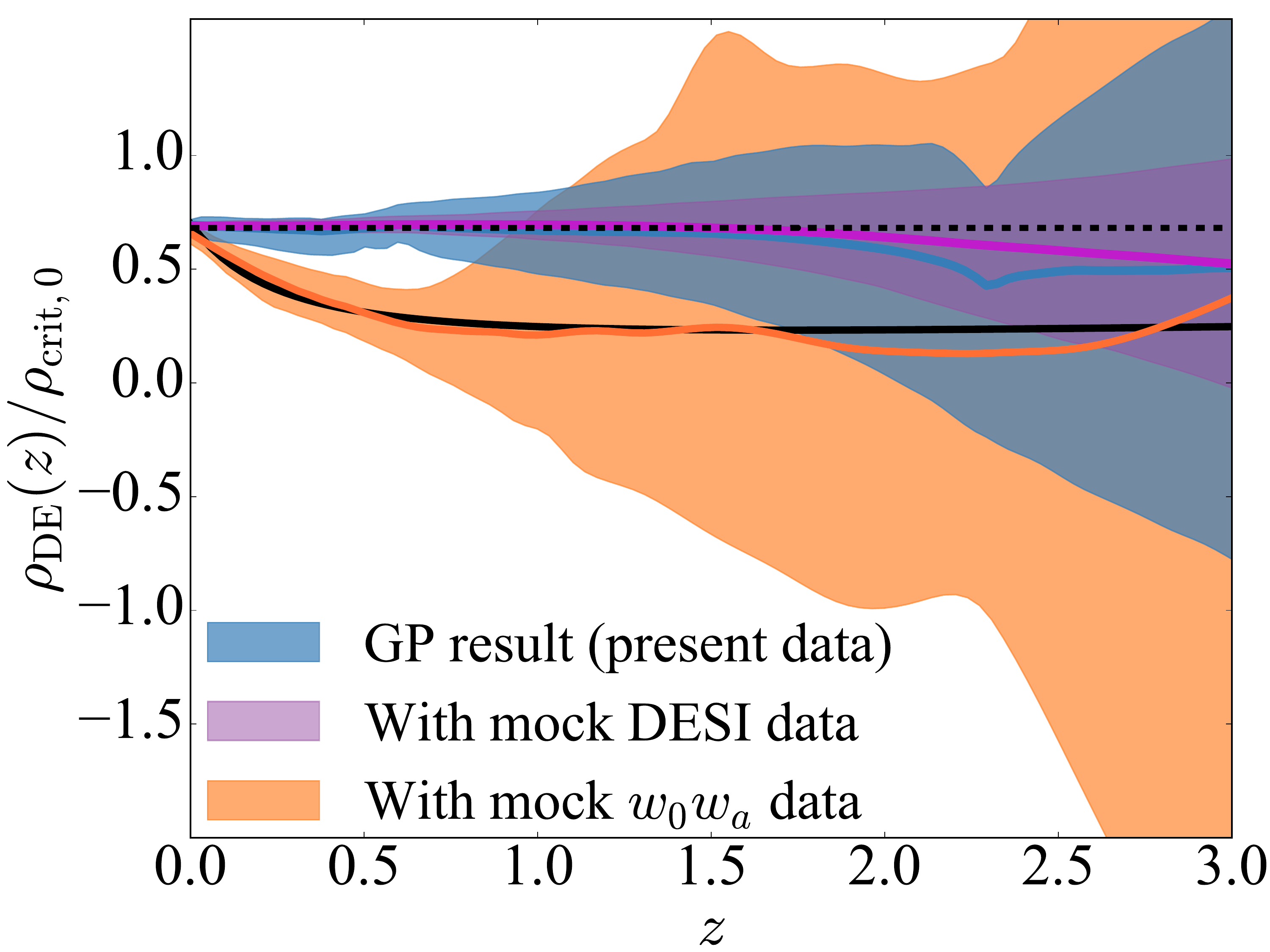}
\vspace{-0.5cm}
\caption{\label{dark_energy} The dark energy density scaled to the present critical density as a function of redshift, as inferred from the expansion histories.  The shaded regions correspond to the posterior probability of the dark energy density at that redshift, bounded at 90\% CL. We consider the full ${\mathsf H0}$-Ly$\alpha$-CMB-SN-LRG dataset combination in blue, and further include forecasted DESI data in purple. In orange, we consider mock data generated from a $w_0 w_a$ cosmology, illustrating that the analysis can recover the dark energy evolution of non-standard cosmologies. The solid black line corresponds to the input $w_0w_a$ model, and the dashed black line is for the $\Lambda$CDM expectation.}
\end{figure}

\begin{subsection}{Dark energy equation of state}

To obtain a sense of the effective dark energy density, we compute the remaining energy density when the matter and radiation energy densities are subtracted from the critical density.  This remaining energy density can be viewed as the dark energy density in a flat cosmology, independent of specific parameterizations for the dark energy equation of state. We define the physical energy densities in matter and radiation in the manner as described above, and compute the effective dark energy density as
\begin{equation}
\rho_{\rm DE}(z)/\rho_{\rm crit,0} = \left(\frac{3H^2(z)}{8\pi G} - \rho_{\rm m}(z) - \rho_{\rm r,fid}(z)\right)\frac{8\pi G}{3 H_0^2} .
\end{equation}
This approach implicitly sets the dark energy density to be zero at the redshift of last scattering. Figure \ref{dark_energy} shows that the inferred dark energy density from current data is consistent with a cosmological constant ($w = -1$) at low redshifts, with a precision of 2\%, 6\%, and 13\% at $z = 0$, 0.5, and 1.0, respectively. As expected, the dark energy constraints successively degrade towards even larger redshifts. The ${\mathsf H0}$ dataset induces some evolution near $z=0$, but it is small compared to the uncertainties in the inferred dark energy density. The inclusion of DESI would reduce the errors by a factor of two to three across redshift, allowing for even more stringent tests of the dark energy equation of state.

In Figure~\ref{dark_energy}, we moreover consider restricting the analysis to mock data generated from a $w_0w_a$ cosmology (same model as in Section~\ref{valsec}), and illustrate that the GP regression is able to recover its dark energy evolution. We find that the size of the errors on the reconstructed dark energy density is sensitive to whether the dark energy is the dominant component of the energy density. When the dark energy density is subdominant to the matter component, the model independent reconstruction has no strong preference for different values of $\rho_{\rm DE}(z)/\rho_{\rm crit,0}$, and allows for large errors. Hence, the mock $w_0 w_a$ cosmology has larger errors than the GP result from present data (despite having the same covariance) as the dark energy in this $w_0 w_a$ cosmology dominates later in time.

\end{subsection}

\end{section}

\begin{section}{Conclusions}\label{sec:conclusion}

We have presented a method for analyzing measurements of the expansion history of the Universe that is independent of cosmological models. To achieve this, we inferred the expansion history using GP regression and showed that the Planck CMB temperature, BOSS luminous red galaxies, BOSS Lyman-$\alpha$, and JLA Type Ia supernova datasets are consistent with one another and with $\Lambda$CDM. The tension between the local Riess~et~al.~(2016) and inferred Planck measurements of the Hubble constant that has been pointed out in the context of $\Lambda$CDM is also apparent in our model-independent analysis. Our analysis did not find evidence that the presence of dark radiation alleviates this tension, leaving open the possibilities for new late-time physics or systematic effects. Beyond $z \simeq 0$, the full combination of datasets constrain the expansion history with a precision of 2\% to the redshift of last scattering, restricting the range of viable non-standard cosmologies.

We derived the growth rate for the fluctuations on sub-horizon scales from the expansion history in a model-independent manner, and showed that it is consistent with the $\Lambda$CDM expectation at the $\lesssim4\%$ level from the present to the matter dominated era. We have not added independent measurements of the growth rate in this work, but doing so in the future will allow more robust tests of deviations from GR. We further constrained the dark energy density with a precision of $2\%$ at $z=0$ and roughly $10\%$ by $z=1$, and found it to be constant across redshift in agreement with the cosmological constant scenario.

We forecasted how the significance of our constraints change with the upcoming DESI experiment. By including DESI in addition to current data, we will be able to be able to improve the constraints on the dark energy density by up to a factor of four, and infer the expansion and growth histories at the percent level from the present to the era of matter domination. This level of precision is encouraging given the model-independent nature of our analysis.

\end{section}

\acknowledgments MK thanks Guillermo Ballesteros for useful discussions. Parts of this research were conducted by the Australian Research Council Centre of Excellence for All-sky Astrophysics (CAASTRO), through project number CE110001020. DK acknowledges support from US Department of Energy award DE-SC0009920. SJ acknowledges support from the Beecroft Trust, STFC, and ERC 693024. The research at UC Irvine was partially supported by National Science Foundation grant PHY-1620638.

\bibliography{master}

\end{document}